\pdfoutput=1
\documentclass[sigconf]{acmart}
\AtBeginDocument{%
  }

\copyrightyear{2026}
\acmYear{2026}
\setcopyright{cc}
\setcctype{by}
\acmConference[IOT 2026]{The 16th International Conference on the Internet of Things}{November 17--20, 2026}{Newcastle Upon Tyne, England, United Kingdom}
\acmBooktitle{The 16th International Conference on the Internet of Things (IOT 2026), November 17--20, 2026, Newcastle Upon Tyne, England, United Kingdom}
\acmDOI{10.1145/3860552.3860597}
\acmISBN{979-8-4007-3095-5/2026/11}
\begin{document}

\title{Continual Instruction Tuning for LLM-based Management of Continuum Resources}

\author{Lanpei Li}
\orcid{0009-0005-4370-1020}
\affiliation{%
  \institution{Institute of Information Science and Technologies ``Alessandro Faedo'', National Research Council (ISTI-CNR)}
  \city{Pisa}
  \country{Italy}}
\affiliation{%
  \department{Department of Computer Science}
  \institution{University of Pisa}
  \city{Pisa}
  \country{Italy}}
\email{lanpeili@cnr.it}

\author{Antonino Vaccarella}
\orcid{0009-0004-0428-540X}
\affiliation{%
  \institution{Institute of Information Science and Technologies ``Alessandro Faedo'', National Research Council (ISTI-CNR)}
  \city{Pisa}
  \country{Italy}}
\affiliation{%
  \department{Department of Computer Science}
  \institution{University of Pisa}
  \city{Pisa}
  \country{Italy}}
\email{antoninovaccarella@cnr.it}

\author{Vincenzo Lomonaco}
\orcid{0000-0001-8308-6599}
\affiliation{%
  \department{Department of AI, Data and Decision Sciences}
  \institution{LUISS University}
  \city{Rome}
  \country{Italy}}
\email{vlomonaco@luiss.it}

\author{Massimo Coppola}
\orcid{0000-0002-7937-4157}
\affiliation{%
  \institution{Institute of Information Science and Technologies ``Alessandro Faedo'', National Research Council (ISTI-CNR)}
  \city{Pisa}
  \country{Italy}}
\email{massimo.coppola@cnr.it}

\renewcommand{\shortauthors}{Li et al.}

\begin{abstract}
Large language models (LLMs) are increasingly explored for resource management across the Cloud--Edge continuum. However, embedding LLMs directly inside latency-sensitive control loops places inference latency, execution jitter, and hallucination risks on the critical path. This study argues for a different use of LLMs: a continually instruction-tuned intent compiler that runs once per management intent or declared context change, above a fast, deterministic tactical controller that handles all low-latency runtime actions. We formulate intent compilation as a continual instruction tuning problem, because management intents arrive sequentially and accumulate while infrastructure and workload contexts drift. Unlike recent LLM-based management frameworks, our design scores the model's outputs by execution, so that catastrophic forgetting can be measured in the executed objective of the underlying controller. On a continuum simulator with a 2B backbone and parameter-efficient adapters, preliminary results show that naive sequential tuning forgets earlier intents in the executed objective, that replay of stored examples nearly prevents this forgetting, and that the tuned compact compiler performs on par with simpler translators while being more robust than a grammar parser to rephrased intents. Code and data are available at \url{https://github.com/lilanpei/continuum-intent-compiler}.
\end{abstract}



\maketitle

\section{Introduction}
\label{sec:intro}
Resource orchestration within the Cloud--Edge continuum requires setting three interdependent control dimensions for distributed multi-stage applications: (i)~\textbf{horizontal elasticity}, sizing worker pools over time; (ii)~\textbf{spatial placement}, mapping pipeline stages across IoT, edge, and cloud nodes; and (iii)~\textbf{capacity allocation}, splitting shared cores between pools. These decisions must satisfy strict Service-Level Objectives (SLOs) governing end-to-end latency and workflow completion while adhering to operational budgets, dynamic energy limits, network bandwidth constraints, and data residency policies. In operational multi-tenant systems, governance requirements are stated as high-level declarative agreements, which intent-based frameworks formulate as natural-language intents~\cite{dzeparoska2023llmbased} (e.g., ``maintain high-priority task deadline compliance above 95\% while strictly prohibiting cloud offloading''). An intent declares \emph{what} is wanted, not \emph{how} to reach it~\cite{clemm2022rfc}; today, systems engineers manually translate it into the low-level operational parameters of the infrastructure controllers~\cite{machado2026orion}.

Our research investigates whether instruction-tuned language models can perform this intent-to-configuration synthesis, and whether they can keep adapting as new kinds of intent arrive and the deployment context drifts, without catastrophic forgetting~\cite{luo2025catastrophic}: the \textbf{Continual Instruction Tuning} (CIT) paradigm, an instance of continual fine-tuning, one of the three directions in which continual learning remains essential in the era of foundation models~\cite{bell2025futurecl}. A model that keeps learning is not a component tuned once and installed but a learning agent that must be designed for adaptation from the start~\cite{quarantiello2026winter}, so where it sits in the platform's control matters as much as what it translates.

We propose two nested control loops: an inner one reacting fast under a deterministic policy, and an outer one carrying the LLM's semantic capabilities, active only when an intent is issued or a context change is declared. The tactical controller (an anticipatory autoscaler or a placement mixed-integer linear program, MILP) acts every epoch in milliseconds on platform state and discrete parameters (\emph{knobs}) alone, while the \textbf{intent compiler}, an instruction-tuned model with a lightweight adapter, runs once per intent, reading the intent together with the current context, and emits a canonical intent and a vector of discrete knobs. A validator checks them before they take effect, and they then stay in force until the next intent or declared context change. Because the compiler is never on the per-epoch path, its answer latency never becomes the platform's control latency, and its supervision arrives only as scarce, expensive labels from scored runs. Against recent LLM-based management systems, the design differs in three ways: the model is \textbf{tuned} on the platform it manages, not only prompted; its output is \textbf{scored} by execution, as its gap to a search reference, not as text; and that score is \textbf{tracked} as intents accumulate and contexts drift.

Preliminary experiments on a continuum simulator with a 2B backbone support this design at reduced scale. Choosing the knobs matters: configurations found by search improve on the schema default by a median of 0.73 in objective value and make the controller act differently. Forgetting is real: tuned stage by stage without protection, the compiler loses 0.23--0.26 of the gain over the default on the intents it learned earlier when new kinds of intent arrive, and 0.67--0.70 under context drift, where the intent text gives no sign that the answer has changed. Forgetting can be reduced: replaying stored examples of earlier stages keeps forgetting at or below 0.07, and the resulting compiler is within 0.05 of one trained on all stages at once. At inference, the tuned compiler needs no simulator rollouts, which are full simulated episodes of the controller's policy scored against the declared objective, and performs on par with a clause-type lookup, a grammar parser, and sixty rollouts of search, and better than retrieval and the zero-shot backbone. When the test intents are rephrased in wordings never seen in training, the compiler keeps most of its accuracy while the grammar parser cannot.

Our contribution is threefold. We propose a two-loop architecture that keeps the tactical controller on the platform's critical path and confines the LLM to an outer loop that runs once per intent. We formulate intent compilation as continual instruction tuning, scored by the executed gap, and test it on two streams that separate the arrival of new kinds of intent from context drift under recurring intents. Finally, we provide preliminary evidence that forgetting under this measure is real, that replay reduces it, and that a tuned compact model is worth its cost against simpler translators.

\section{Related work}
\label{sec:related}
Human operators can declare structured requirements --- intent-based networking supplies a vocabulary and lifecycle~\cite{clemm2022rfc}, and elasticity SLOs are first-class objects in dedicated controllers~\cite{pusztai2021polaris} --- but in such controllers the mapping from declared objective to configuration is fixed by hand, not learned from open-ended input.

Transforming an intent into a concrete configuration with an LLM is an active line of research.
Policies and configurations are generated from natural language~\cite{dzeparoska2023llmbased}, with accuracy that depends on the task, the model, and the prompt~\cite{wang2024netconfeval}.
Intent-driven controllers now reach the compute continuum~\cite{akbari2025intentcontinuum}, and agent compilers for radio infrastructure separate timescales in a way similar to ours: an LLM on a second-to-minute scale above a millisecond-scale control loop~\cite{zheng2026agentcompiler}.
Among these LLM-based intent-to-configuration systems, some already screen the output before it acts, through an intent contract~\cite{bimo2026caif}, a constraint check over a verifiable intermediate representation~\cite{zheng2026agentcompiler}, or an LLM check and a digital-twin test~\cite{dzeparoska2023llmbased}; elsewhere it is applied as returned~\cite{akbari2025intentcontinuum}.
All of them use an LLM whose weights stay fixed once deployed, from GPT-4o down to a 4B instruct model; where there is a gate, it only ever validates one output at a time---the configuration just returned. We instead continually tune a compact adapter offline, so our gate must also decide whether a candidate adapter may replace the active one, admitting it only if it still performs on the intents the active adapter already learned. A recent taxonomy~\cite{vaccarella2026smartadaptivecomputingcontinuum} classifies six systems by whether execution feedback returns to the LLM after agents act. None of these systems uses execution outcomes to update the LLM's weights.

Continual learning studies how a model can learn from a sequence of tasks over time without losing what it has already learned; the risk is catastrophic forgetting: learning a new task degrades performance on what was learned before~\cite{luo2025catastrophic}.

\section{Problem formulation}
\label{sec:problem}
From an engineering viewpoint, when an LLM is placed in a controller's feedback loop, its properties create a dependency cycle. The size and type of the model dictate its capabilities, but also how fast it can answer and its hardware needs. The model latency is a lower bound on the control latency of the whole system, and control latency dictates how quickly the system can react and how often the model can learn from feedback, thus limiting its overall control capability. We break this cycle by giving the model a different job: it produces a configuration once per intent and steps aside while a fast, deterministic controller acts at every epoch under it. Its feedback is \textbf{late} and \textbf{mediated}, the measured outcome of a whole run rather than a per-epoch reward, and because intents keep arriving, these outcomes accumulate as a stream of labelled examples; learning from that stream without losing earlier competence is the continual instruction tuning problem.

\subsection{Platform, controller, and knobs}\label{sec:controller}
We define a continuum platform with four layers (space, IoT, edge, cloud), where space and IoT nodes host the data sources, and edge and cloud are the compute layers.  Applications are directed acyclic graphs (DAGs) of pipeline stages. Time advances in \textbf{control epochs} (one second by default). At epoch $e$ the platform exposes a state $s_e$: the current placement, the per-stage queues and their ages, and the workers of each pool. A \textbf{tactical controller} $\pi_{\boldsymbol{\theta}}$, parameterised by a vector $\boldsymbol{\theta}\in\Theta$, maps $s_e$ to a feasible action $a_e$: a placement of stages on nodes and a size for each worker pool. When pools share nodes, their sizes split the common core budget. The controller executes deterministically at each epoch and applies only actions the platform can carry out: pool sizes stay within their floors and caps, and a placement that would exceed a node's capacity is never applied; the controller falls back to one that fits, at worst the current placement. In our experiments $\pi_{\boldsymbol{\theta}}$ is an anticipatory autoscaler that pre-warms and drains pools around scheduled bursts on the elastic and two-pool variants of Section~\ref{sec:substrate}, and a MILP that places stages under capacity, residency, link, and energy constraints on the placement variant; Section~\ref{sec:knobs} lists their knobs. The controller is conventional software: it is never trained or modified, and the only thing the compiler changes is $\boldsymbol{\theta}$.

Each entry of $\boldsymbol{\theta}$ is a \textbf{knob} that a human engineer would set when translating a service-level agreement: an objective weight, a per-class deadline, a scaling target, a pre-warm lead time (how early workers start before a burst), a pool floor or cap, a layer price weight, a residency rule, or a hard cap on data crossing into the cloud or on energy.  Every knob is \textbf{discrete}, with three to five allowed values, so $\Theta$ is a finite set. The list of knobs with their types, sets of allowed values, and defaults is the \textbf{schema}; it establishes $\Theta$ and the serialisation of $\boldsymbol{\theta}$. With the clause grammar of Section~\ref{sec:intents}, the schema forms the complete target language of the compiler.

\begin{table*}[t]
    \caption{The eight intent families and their clauses ($\bullet$ required, $\circ$ optional), grouped by the scenario variant defined in Section~\ref{sec:substrate}.}
  \label{tab:clauses}
  \centering
  \footnotesize
  \setlength{\tabcolsep}{4pt}
  \begin{tabular}{@{}lcccccccccc@{}}
    \toprule
    & \multicolumn{9}{c}{Intent clause type and its forms} \\
    \cmidrule(lr){2-10}
    & latency & priority & \multicolumn{2}{c}{budget} & \multicolumn{2}{c}{energy} & residency & link & throughput \\
    \cmidrule(lr){2-2}\cmidrule(lr){3-3}\cmidrule(lr){4-5}\cmidrule(lr){6-7}\cmidrule(lr){8-8}\cmidrule(lr){9-9}\cmidrule(lr){10-10}
    Family & latency (deadline classes) & class first & cost weight & capacity cap & energy weight & power cap & allowed layers & link budget & completion floor \\
    \midrule
    \multicolumn{10}{@{}l}{\emph{Elastic variant: deciding how many workers to run over time}} \\
    budget-elastic      & low       & & $\bullet$ & & $\circ$ & & & & $\circ$ \\
    bill-elastic        & low       & & $\circ$ & $\bullet$ & & & & & \\
    \midrule
    \multicolumn{10}{@{}l}{\emph{Placement variant: deciding where stages run}} \\
    layer-placement      & high, low & & $\bullet$ & & & & & & \\
    link-placement      & high, low & & $\bullet$ & & & & & $\bullet$ & \\
    residency-placement & high, low & & $\bullet$ & & & & $\bullet$ & $\circ$ & \\
    \midrule
    \multicolumn{10}{@{}l}{\emph{Two-pool variant: deciding how a shared core budget is split between pools}} \\
    priority-two-pool   & high, low & $\bullet$ & $\circ$ & & & & & & \\
    budget-two-pool     & high, low & & $\bullet$ & & & & & & \\
    bill-two-pool       & high, low & $\bullet$ & & $\bullet$ & & & & & \\
    \bottomrule
  \end{tabular}
    \par\noindent\parbox{0.975\linewidth}{\emph{\textbf{Note:} budget and energy each have a soft (weight) and a hard (cap) form; an empty entry means the clause is not yet used in experiments, not disallowed by the grammar.
    }}
  \Description{Rows are the eight intent families, and columns are the nine clause forms of the intent grammar. Each cell marks a clause form as required, optional, or unused for that family, except in the latency column, whose cells name the task classes for which the family's intents state a deadline.}
\end{table*}

\subsection{Intents, contexts, and the declared objective}\label{sec:intents}
\textbf{Intent and grammar.} The LLM reads an intent and a context, and produces an intermediate form that a human can audit before the parameters take effect. An \textbf{intent} $\iota$ is a natural-language sentence or short paragraph stating what the platform should achieve.  Its \textbf{canonical intent} $\kappa$ is a list of typed clauses, each with its parameters.  The clause types (Table~\ref{tab:clauses}) are latency (a deadline for one class of tasks), priority (one class guaranteed an on-time rate, the other served best effort), budget (priced capacity, as a cost weight or a hard cap on the share used), energy (a weight or a cap on mean dynamic power), residency (which layers may run stages), link (a cap on the data that may cross into the cloud), and throughput (a floor on the fraction of tasks completed).  For example, a sentence such as ``at most 64\,MB may cross into the cloud during the run'' is one link clause. A \textbf{soft} clause states a preference (e.g., over cost or energy) and scales its relative priority via an objective weight; a \textbf{hard} clause (e.g., a cap, floor, or residency rule) represents a mandatory constraint whose violation yields an infeasible output rather than a suboptimal one. These clause types, in the nine forms of Table~\ref{tab:clauses} and each with a fixed set of parameter values, make up the intent \textbf{grammar}; each intent family fixes which clauses its intents combine, and a family is admitted only when the clause it adds can bind. Finally, clauses are stated at the level at which a platform is managed, and the choice of nodes is left to the controller: cost is defined per layer, residency names layers, and energy is a single platform-wide total of dynamic power.

\textbf{Context.} An intent alone is insufficient, as the same high-level specification requires different parameter configurations across different platforms. We therefore pair the intent with a deployment \textbf{context} $c$, a structured snapshot of the system at the moment an intent is issued. It encompasses the application DAG, layer capacities and pricing, workload intensity and arrival profiles, and a one-line summary of a baseline rollout under default parameters.

\textbf{Declared objective.} The \textbf{declared objective} $J_{\iota}(\boldsymbol{\theta};c)$ evaluates the executed score of a rollout against the criteria specified in $\kappa$; we call its value the \textbf{objective value} of $\boldsymbol{\theta}$ in context $c$ for intent $\iota$ with canonical intent $\kappa$. It builds on the completion-aware score of ContinuumBench~\cite{li2026continuumbench}: each soft clause in $\kappa$ sets an objective weight, each hard clause introduces a violation penalty, and every other term keeps its default weight. If the controller cannot execute a rollout, that rollout receives a fixed, negative floor score. This score is evaluated by rolling out policy $\pi_{\boldsymbol{\theta}}$ in context $c$ over a full episode and averaging across simulator seeds.

\textbf{Fixed and free knobs.} Knobs are split by whether they define \textbf{what} the intent wants or \textbf{how} the controller achieves it. A knob is \textbf{fixed} if it defines the target goal: the fixed knobs are the objective's weights, set by soft clauses, and its per-class deadlines, set by latency clauses, with defaults for any term the intent omits. Allowing the compiler to choose these would let it choose the very objective it is scored against. Conversely, a knob is \textbf{free} if it governs operational mechanisms. An intent never names a pre-warm lead or a layer price weight, so the compiler must infer such knobs from the intent and the deployment context together. Only the residency and link clauses set a free knob directly; other caps and floors are met through free knobs the compiler chooses. Accordingly, we partition the parameters as
$\boldsymbol{\theta}=(\boldsymbol{\theta}_{\mathrm{fix}},\boldsymbol{\theta}_{\mathrm{free}})$, with $\boldsymbol{\theta}_{\mathrm{fix}}=h(\kappa)$ set by the canonical intent.

\subsection{Intent compilation as a supervised task}\label{sec:intentComp}
The free knobs encode quantitative values unlikely to appear in a general pretraining corpus, such as how long before a burst the worker pools must be pre-warmed, or how heavily the placement should weigh edge against cloud prices. Without observing actual rollouts, a model can only guess these values. Large hosted models might guess better, at a monetary cost, with an external dependency, and with operational data leaving the platform.

We therefore fine-tune a compact, local model to predict the free knobs. Because there is no analytical ground truth for them, the label comes from a search process.

An \textbf{intent compiler} is a function
\begin{equation}
  f_\phi:(\iota,c)\mapsto(\widehat{\kappa},\widehat{\boldsymbol{\theta}}),
  \label{eq:compiler}
\end{equation}
implemented as a pretrained LLM backbone, kept \emph{frozen} (its weights are never updated), plus a lightweight trainable adapter $\phi$, a form of parameter-efficient fine-tuning (PEFT). A downstream \textbf{validator} $V$ first verifies syntactic correctness, accepting the output only if it conforms to the target schema and every knob falls within its permitted domain. Operational hard constraints encoded in $\widehat{\boldsymbol{\theta}}$ (e.g., residency rules, link budgets, and energy caps) are enforced by the controller, which flags the configuration as infeasible if the platform cannot satisfy them. The end-to-end execution contract is:
\begin{equation}
  (\iota,c)
  \xrightarrow{f_\phi}(\widehat{\kappa},\widehat{\boldsymbol{\theta}})
  \xrightarrow{V}\pi_{\widehat{\boldsymbol{\theta}}}
  \xrightarrow{\text{rollout}}
  J_{\iota}(\widehat{\boldsymbol{\theta}};c).
  \label{eq:interface}
\end{equation}
\paragraph{Reference labels.}
In offline training mode (see Section \ref{sec:opmodes}) the supervisory target for the compiler is the best configuration discovered by bounded search on the simulator:
\begin{equation}
  \boldsymbol{\theta}^\star(\iota,c)\approx
  \arg\max_{\boldsymbol{\theta}\in\Theta,\;\boldsymbol{\theta}_{\mathrm{fix}}=h(\kappa)}
  J_{\iota}(\boldsymbol{\theta};c),
  \label{eq:reference}
\end{equation}
where a greedy search changes one or two free knobs at a time, and a few finalists are re-scored on additional seeds. Because a single reference label costs a search of one to two thousand rollout evaluations, most of them served from a cache shared across the labels of one context, labels are expensive and scarce.

\paragraph{Training objective.} We cast compilation in a standard instruction-following format: the \textbf{instruction} is the intent text $\iota$, the \textbf{input} is the serialised context $c$, and the supervisory \textbf{target} is the serialised pair $(\kappa,\boldsymbol{\theta}^\star_{\mathrm{free}})$. Fixed knobs are excluded from generation because they are deterministically derived via $\boldsymbol{\theta}_{\mathrm{fix}} = h(\kappa)$. The adapter is trained with the usual token-level cross-entropy on the target.

\paragraph{Measuring a compiled configuration.}
We score the output by what happens when it is executed, since two different parameter vectors can be operationally equivalent and a near-identical one can be infeasible. The \textbf{executed gap} we measure is defined as
\begin{equation}
  g(\widehat{\boldsymbol{\theta}};\iota,c)=
  \frac{J_{\iota}(\boldsymbol{\theta}^\star;c)-J_{\iota}(\widehat{\boldsymbol{\theta}};c)}
       {\max\!\big(\delta,\;J_{\iota}(\boldsymbol{\theta}^\star;c)-
        J_{\iota}(\boldsymbol{\theta}_{\mathrm{def}};c)\big)},
  \label{eq:gap}
\end{equation}
where $\boldsymbol{\theta}_{\mathrm{def}}$ denotes the default configuration under the schema, and $\delta = 0.05$ serves as a floor to stabilise the denominator when the default is near-optimal; the default's own gap is then below 1. The gap is clipped to $[0,1]$. Intuitively, a gap of $0$ indicates performance matching the search reference $\boldsymbol{\theta}^\star$, whereas a gap of $1$ signifies performance no better than the default baseline; rejected or unparsable outputs are assigned a gap of $1$.

We define three scores for every example, which separate reading the intent from choosing the knobs.  The \textbf{parse} score asks whether the intent was read correctly: it checks for an exact match between the predicted canonical intent $\widehat{\kappa}$ and the gold $\kappa$. The \textbf{solve} score asks how good the knobs are assuming the intent is parsed correctly: it is the \textbf{gap}~\eqref{eq:gap} obtained when the gold $\kappa$ is supplied and only $\boldsymbol{\theta}$ has to be produced. The \textbf{end-to-end} score is the gap of the full pipeline, parsing $\iota$ to $\kappa$ and choosing $\boldsymbol{\theta}$ together, and it is our final, long-term target.

\subsection{Continual instruction tuning of the compiler}
\label{sec:cit}

The compiler must also keep learning after deployment, for three reasons. First, new kinds of intent appear while the earlier ones stay in use. Second, prices, capacities, and workloads drift, so the best knobs $\boldsymbol{\theta}^\star$ for an unchanged intent change although its text does not. Third, every label costs a search, so new data arrive in small batches, and retraining on all past data at every update would exceed the platform's compute and storage budget.

Learning proceeds in $K$ \textbf{stages}. At stage $k$, the compiler receives a batch $\mathcal{D}_k$ of labelled examples sampled from an intent--context distribution $\mathcal{P}_k$. The adapter $\phi$ is updated on $\mathcal{D}_k$, augmented by an optional replay memory $\mathcal{M}_k$ storing at most $m$ examples per preceding stage:
\begin{equation}
  \phi_k = U(\phi_{k-1},\,\mathcal{D}_k,\,\mathcal{M}_k),
  \qquad |\mathcal{M}_k|\le m\,(k-1),
  \label{eq:update}
\end{equation}
where $U$ denotes the continual update rule under study, and $m$ the per-stage memory quota ($m=0$ for memory-free methods). All other components---the LLM backbone, schema, validator, tactical controller, and evaluation pipeline---stay unchanged throughout. At test time, the compiler operates task-agnostically: it receives only $(\iota,c)$ with no stage identifier, and as in CIT benchmarks~\cite{zhang2023citb,wang2023trace}, it is evaluated on every stage learned so far.

A continual compiler can fail in two ways: it can perform poorly on the tasks it has been taught, and it can lose earlier competence while learning new ones. Let $g_{k,j}$ be the mean end-to-end gap of $f_{\phi_k}$ on the held-out test set of stage $j\le k$. We report the \textbf{Average Gap (AG)} after the last stage, $\mathrm{AG}=\frac1K\sum_{j=1}^{K}g_{K,j}$, the compiler's competence over everything it has been taught, and \textbf{Executed Forgetting (F)}, $F=\frac1{K-1}\sum_{j<K}\big(g_{K,j}-\min_{k\ge j}g_{k,j}\big)$, the average growth of each earlier stage's gap since its best value. Both AG and $F$ are measured in executed objective, not text similarity.

\section{Architecture}
\label{sec:architecture}

\begin{figure*}[t]
  \centering
  \includegraphics[trim=0mm 0mm 0mm 0mm, width=0.85\textwidth]{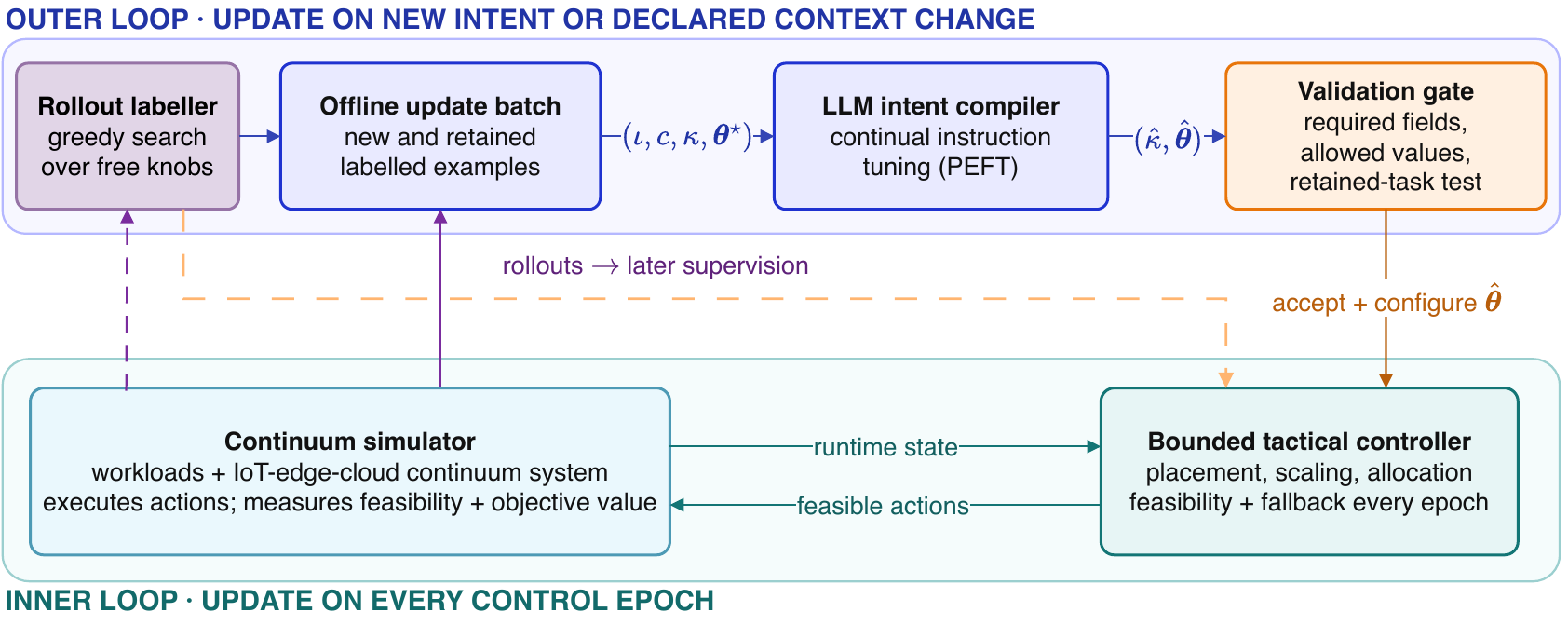}
  \caption{The two nested control loops. Top: the outer loop, in which the compiler, tuned offline on the update batch, turns an intent and its context into $(\widehat{\kappa},\widehat{\boldsymbol{\theta}})$ and the validation gate admits them to the tactical controller or rejects them. Bottom: the inner loop, in which the tactical controller acts at every epoch on the continuum simulator. Measured outcomes return only to the offline adapter update. Dashed arrows depict exchanges that happen only during offline label search.}
  \Description{The outer loop contains a rollout labeller, an offline update batch of labelled examples, a continually instruction-tuned LLM intent compiler, and a validation gate.  An accepted output configures the tactical controller.  In the inner loop, the controller exchanges runtime state and feasible actions with the continuum simulator, which represents workloads and IoT, edge, and cloud resources.  Measured outcomes return to later offline updates.}
  \label{fig:architecture}
\end{figure*}

Figure~\ref{fig:architecture} shows the architecture that realises the execution contract~\eqref{eq:interface}. The high-frequency inner loop runs the tactical controller at every epoch with millisecond latency. The low-frequency intent compilation occurs in the outer loop, \textbf{triggered} either by a new intent or by a declared context change.

\subsection{Operation modes}\label{sec:opmodes}
System operation is decoupled into two phases, online inference and offline adaptation.

\textbf{At inference time}, on either trigger, the compiler reads the intent and the serialised context, then generates and emits $\widehat{\kappa}$ and $\widehat{\boldsymbol{\theta}}$. If they are accepted by the validator $V$, $\widehat{\boldsymbol{\theta}}$ is written once into the controller's configuration, and the inner loop runs from it at every epoch until the next event.
No label is needed and no parameter search is performed (see Section~\ref{sec:intentComp} and Eq.~\eqref{eq:reference}), so intent compilation needs no rollouts. On one NVIDIA A100 GPU, the 2B backbone takes about 0.6\,s per intent when decoded in batches.

\textbf{Offline}, the intents issued since the last update and their evaluated outcomes are collected; each new intent--context pair is labelled by the bounded search~\eqref{eq:reference} on the simulator; the pairs form the next stage's batch; a candidate adapter is tuned on it under the memory quota of~\eqref{eq:update}, its outputs on retained intents and contexts are run through the controller, and it replaces the active adapter only if, on those examples, its mean gap exceeds that of the active adapter by at most $\epsilon$, with $\epsilon$ just above the gap's noise across training seeds and decodings (up to 0.06). In our experiments every candidate is accepted, so that forgetting can be measured. The simulator is therefore used only offline, to produce labels and to test candidate adapters before they go live; at inference the compiler stands in for it.  A label is only as good as the simulator's match to the platform; in this study the simulator is also the platform, so that mismatch is not measured.

\subsection{Validation gate and adapter updates}\label{sec:gate}
The validator $V$ rejects an invalid output instead of repairing it: snapping an invalid value to the nearest allowed one would silently replace a wrong answer with another one, whereas rejection keeps the last accepted configuration active and makes the failure visible. The gate catches only what it encodes: a constraint the schema omits, or a poor but feasible $\widehat{\boldsymbol{\theta}}$, passes it.

In our experiments, only the adapter (IA$^3$~\cite{liu2022ia3}, which learns vectors that rescale the backbone's inner activations) is trained on each arriving batch, as in~\eqref{eq:update}. It is a small fraction of a per cent of the backbone's parameters, so an update is a small file and a rollback is a file swap.

\section{Experimental setup}
\label{sec:setup}\label{sec:substrate}
Experiments are conducted on ContinuumBench~\cite{li2026continuumbench}, a discrete-time simulator modelling distributed stream-processing workflows and Earth-observation pipelines deployed across the four layers of Section~\ref{sec:controller}. Workloads are task streams with Poisson arrivals whose rate is either constant or follows a schedule of rate segments over time; in our experiments, capacity is priced per layer and energy is metered per node. The experiments use its Earth-observation pipeline, fed by a space stream and an IoT stream, on three \textbf{scenario configuration variants}; in two of them the IoT stream arrives in short periodic bursts, and the scheduled burst is what the anticipatory autoscaler pre-warms for.  Each variant's base configuration fixes the application, infrastructure, and workload so that one control lever decides the outcome.  The \textbf{elastic} variant exercises scaling: a bottleneck pool must be sized over time against the bursts, and each new worker takes 8\,s to start.  The \textbf{placement} variant exercises where stages run: a DAG needing 32 cores must be split across a 24-core cloud node and two 12-core edge nodes, with a 150\,ms, 40\,Mbps link between edge and cloud, so it fits on neither side alone.  The \textbf{two-pool} variant exercises allocation: two pools share the 44 cores of the edge and cloud nodes and room for one worker on the space node, too little for both at peak.

\subsection{Tactical controllers and their knobs}\label{sec:knobs}

The thirteen free knobs are the settings of the two controllers of Section~\ref{sec:controller}. The placement MILP maps stages to nodes; its five knobs are the edge and cloud layer price weights, the residency rule, the link budget, and the energy cap.  The anticipatory autoscaler sizes each worker pool at every epoch: it pre-warms a pool ahead of a scheduled burst, holds it for a drain time, and falls back to queue-driven scaling for load the schedule did not predict; its eight knobs are the pre-warm lead and targets, the drain time, the pool floors and cap, and the reactive queue target.  The six fixed knobs set the per-class deadlines and the objective's weights, and the cost and energy weights also enter the MILP objective.  The schema has nineteen knobs in all.

\subsection{Intent and instruction design}
We instantiated each of the nine clause forms of Table~\ref{tab:clauses} with handwritten sentence templates and expanded them with a large language model into paraphrases, giving 104 training wordings (37 templates and 67 paraphrases) and 38 held-out rewordings. To keep the test wordings distinct from the training ones, a word-overlap filter keeps the Jaccard similarity between each held-out phrasing and every training sentence below 0.5. Wordings are short sentences that state a class, a limit, or a location. To test robustness to rewording, each test intent is evaluated under both training and held-out wordings.

The clauses form the eight \textbf{intent families} of Table~\ref{tab:clauses}, each a combination of clauses on one variant. Every family carries a latency clause because a deadline is what every other clause trades against, and each adds one clause type or one conflict on a variant where it changes the outcome, so the eight families are eight distinct things to learn.

\subsection{Input streams}
\label{sec:streams}
Between learning stages the distribution of intent--context pairs shifts, and it can shift on either side: in the intents that are issued, or in the contexts they are issued for.

Text benchmarks for continual instruction tuning typically employ an input stream in which every stage brings a new task with new instructions~\cite{zhang2023citb,wang2023trace}.
Such a stream cannot separate retention from recognition: a classifier that recognises the task from the instruction text can route each input to the right adapter and never have to forget. It also cannot say anything about the case in which the instructions stay the same while the right answer changes. We therefore define two distinct input streams and keep these two kinds of changes apart while testing.

\textbf{Stream I: intent accumulation.}
Each of its eight learning stages introduces one new intent family, while each variant's contexts are drawn from the same ranges of prices, capacities, and workloads at every stage.
The intents of the earlier families remain in use, and the compiler is tested on them. Because most families differ in their clauses, the instruction text mostly identifies the stage. We call such stages \textbf{instruction-separable}.

\textbf{Stream II: context drift under recurring intents.}
Here the intents recur but the deployment changes: the same families return while the ranges from which their contexts are drawn shift between stages, which we call \textbf{context drift}. Contexts come from two \textbf{context settings}: the first makes the edge smaller and cheaper and the cloud more expensive than the second does. Prices differ by a factor of two to four between them. Each stage is one of four families (budget-elastic, layer-placement, link-placement, priority-two-pool) under one setting, so Stream~II also has eight stages, and $F$ averages over seven earlier stages in both streams. Each stage draws its contexts, and hence its test set, from its own setting only. The search reference $\boldsymbol{\theta}^\star(\iota,c)$ therefore changes while the wording of $\iota$ does not, so the compiler must read the change off the context.  No function of $\iota$ alone can identify the stage, and we call such stages \textbf{instruction-inseparable}.

\textbf{Stage orders.} Since forgetting is sensitive to stage ordering, we evaluate each stream under two orderings, Order~A and Order~B. In Stream~I, Order~A alternates across variants, so consecutive families share few knobs. Order~B groups families by variant, letting consecutive stages compete over the same knobs with conflicting targets. In Stream~II, which evaluates the families under two context settings, Order~A introduces all families in the first setting before the second (creating a single context drift event), whereas Order~B alternates settings family by family (continuous drift). Each update rule is evaluated over all four combinations, with results reported as the range across the two orders.

\subsection{Models and methods}
The backbone is Gemma-2B-it~\cite{gemmateam2024gemma} with IA$^3$ on all decoder blocks (Adam, learning rate $10^{-4}$, batch 16). Two update rules are studied: plain sequential tuning of the adapter, with no protection against forgetting, and standard experience replay under the memory quota of~\eqref{eq:update}, caching 50 exemplars per earlier stage and adding them to every later stage's training data. As an upper bound, the \textbf{joint ceiling} trains one adapter on all stages of a stream at once, with about the same total number of optimiser steps as the whole sequential run. Separately from the update rules, the tuned compiler itself is compared against other ways of turning an intent and a context into $\boldsymbol{\theta}$. These translation baselines (see Table \ref{tab:baselines}) use the same schema and the same rollout scoring as the compiler, and they fall into four groups, each answering one question. \textbf{Group 1} is the floor: the schema default. \textbf{Group 2} is given the correct canonical intent, so its gap is the solve score; it asks whether choosing the knobs needs a generative model once the parse is known. A \emph{clause-type lookup} returns one stored $\boldsymbol{\theta}$ per intent family and combination of clause types, and ignores the context. A \emph{regression head} predicts each free knob from the frozen backbone's embedding of the prompt (intent text and context), taking the canonical intent and fixed knobs from the gold parse. \emph{Search at inference} runs a greedy single-knob search on the simulator at test time, with a budget of 10, 30, or 60 rollouts. \textbf{Group 3} is given only the intent text and the context and runs no rollouts, so its gap is the end-to-end score. The \emph{zero-shot} backbone tests whether tuning is needed at all. A \emph{grammar parser plus rules} matches the text against the training wordings and then applies the lookup, and \emph{retrieval} returns the label of the most similar training example; both test whether a tuned model is needed. The \emph{joint ceiling} and the replay-tuned compiler are the tuned models. \textbf{Group 4} starts from the replay-tuned compiler's output and spends sixty rollouts of search; it asks whether search can replace or complement the compiler. \emph{Parse + search from the default} keeps only the compiler's parse, resets the free knobs to the schema defaults, and searches from there. \emph{Answer + search from that answer} starts the search at the full configuration the compiler emitted.

\subsection{Success criteria}
The experiments test four criteria:

\textbf{The knobs matter}: the search reference $\boldsymbol{\theta}^\star$ must substantially outperform the schema default $\boldsymbol{\theta}_{\mathrm{def}}$ in objective value and induce distinct controller behaviour (e.g., different placements or worker counts); if default parameters perform competitively, compilation provides no practical benefit.

\textbf{Compiling is not a lookup}: the tuned compiler must achieve a lower gap than the two model-free shortcuts, the clause-type lookup and retrieval, on contexts unseen during training; if a shortcut suffices, the mapping is trivial and no model is needed.

\textbf{Forgetting is real}: when the compiler is tuned stage after stage with no protection against forgetting, its forgetting $F$ must erode a clear share of the improvement that the search reference offers over the default.

\textbf{Forgetting can be reduced}: at least one update rule must cut $F$ substantially below that of unprotected tuning.

\section{Preliminary experimental results}
\label{sec:exp}
We ran the protocol of Section~\ref{sec:setup} on one backbone, with all 60 contexts per stage (36/12/12 train/validation/test) and three training seeds for every trained model except the Stream~II joint ceiling and the regression head, which have one. In deployment the validator rejects an invalid output (Section~\ref{sec:gate}). For measurement, rejection would score every invalid output at gap 1 and hide how close it came to a usable answer. When scoring, we therefore repair it (unknown knobs dropped, values outside the allowed set snapped to the nearest allowed value, missing knobs set to their defaults) and execute the result: repair is a scoring convention, not part of the system. Repair can only lower a gap, so the gaps reported below are lower bounds on what the deployed validator would charge, most of all for sequential tuning, whose outputs are invalid far more often than replay's. Table~\ref{tab:exp} reports the two update rules and Table~\ref{tab:baselines} the translation baselines.

\begin{table}[t]
  \caption{Preliminary results on the 2B backbone (eight stages per stream, two orders; lower is better).}
  \label{tab:exp}
  \centering
  \small
  \begin{tabular}{@{}lcccc@{}}
    \toprule
    & \multicolumn{2}{c}{Stream I} & \multicolumn{2}{c}{Stream II} \\
    \cmidrule(lr){2-3}\cmidrule(lr){4-5}
    Method & AG & $F$ & AG & $F$ \\
    \midrule
    Sequential IA$^3$ & 0.51--0.53 & 0.23--0.26 & 0.74--0.75 & 0.67--0.70 \\
    Replay (50/stage) & 0.29--0.33 & 0.03 & 0.17--0.18 & 0.06--0.07 \\
    \midrule
    Clause-type lookup & 0.31 & -- & $\ast$ & -- \\
    Retrieval & 0.41 & -- & $\ast$ & -- \\
    Joint ceiling & 0.28 & -- & 0.15 & -- \\
    \bottomrule
  \end{tabular}
   \par\parbox{0.94\linewidth}{\footnotesize Each stream is run in both stage orders. A range gives the two
   per-order values.\newline (--) denotes an inapplicable metric; ($\ast$) not evaluated on that stream's test set.}
  \Description{The table lists the average gap (AG) and forgetting (F) of sequential tuning and of replay on Streams I and II, with the clause-type lookup, retrieval, and the joint ceiling for comparison. Replay forgets far less than sequential tuning, and its average gap comes close to that of the joint ceiling.}
\end{table}

\textbf{Knobs drive performance, but simple lookups suffice on certain families.}
The first criterion holds: across six representative intents, one for each constraint that can bind (capacity, budget, energy, link, priority, residency), the parameters found by search improve on the default by a median of 0.73 in objective value, and the references of any two of these intents set different worker counts or placements, so the knobs change what the controller does, not only how a run is scored; choosing the knobs therefore matters.  The second criterion holds only on some families.  The clause-type lookup reaches a mean gap of 0.31 on Stream~I when given the correct canonical intent, close to the 0.28 of the joint ceiling, the best the tuned compiler reaches here. The 0.31 is a mean over eight families, whose gaps range from 0.08 to 0.60. On the budget-elastic and budget-two-pool families, the lookup's gap is only 0.08 and 0.17, so one stored vector per combination of clause types already serves well there. On the priority-two-pool and layer-placement families, the lookup loses 0.45 and 0.60 of the improvement that the search reference offers over the default: for these intents the best free knobs change with the prices and capacities of the context, a stored vector cannot follow them, and a model that reads the context is needed.

\begin{table}[t]
  \caption{The tuned compiler against the translation baselines on the Stream~I test set. Each entry is the mean gap~\eqref{eq:gap} over the test intents (lower is better).}
  \label{tab:baselines}
  \centering
  \footnotesize
  \setlength{\tabcolsep}{4pt}
  \begin{tabular}{@{}lccc@{}}
    \toprule
    Method & Rollouts & Seen & Reworded \\
    \midrule
    \multicolumn{4}{@{}l}{\emph{Group 1: the floor}} \\
    Default $\boldsymbol{\theta}$ (no knob chosen)              & 0  & 0.93 & 0.93 \\
    \midrule
    \multicolumn{4}{@{}l}{\emph{Group 2: given the correct canonical intent (solve score)}} \\
    Clause-type lookup                              & 0  & 0.31 & -- \\
    Regression head (frozen embeddings)             & 0  & 0.29 & 0.41 \\
    Search at inference                             & 60 & 0.29 & -- \\
    \midrule
    \multicolumn{4}{@{}l}{\emph{Group 3: given only the intent text and the context, no rollouts (end-to-end score)}} \\
    Zero-shot Gemma-2B-it                           & 0  & 0.92 & -- \\
    Grammar parser + rules                          & 0  & 0.31 & 0.84 \\
    Retrieval (intent + context)                    & 0  & 0.41 & -- \\
    Joint ceiling                             & 0  & 0.28 & 0.33 \\
    IA$^3$+Replay                 & 0  & 0.29--0.33 & 0.37--0.38 \\
    \midrule
    \multicolumn{4}{@{}l}{\emph{Group 4: given only the intent text and the context, then sixty rollouts of search}} \\
    IA$^3$+Replay, parse + search from the default     & 60 & 0.62 & -- \\
    IA$^3$+Replay, answer + search from that answer    & 60 & 0.21 & -- \\
    \bottomrule
  \end{tabular}
    \par\parbox{\linewidth}{\footnotesize Tuned models and Group~4: means over three training seeds; IA$^3$+Replay gives the two stage orders, as in Table~\ref{tab:exp}, and Group~4 starts from Order~A. Search at inference scores each candidate on one simulator seed, then re-scores its best candidate and the default on three (about six more rollouts); Group~4 scores every candidate on three. Seen: training wordings; Reworded: held-out paraphrases; (--) not evaluated.}
  \Description{The methods are grouped by what they are given; for each, the table shows the number of rollouts spent per intent and the gap on training wordings and, where tested, on rewordings. The joint ceiling and the replay-tuned compiler are on par with the strongest baselines on training wordings and far better than the grammar parser on rewordings. Search started from the compiler's answer, with sixty rollouts, has the lowest gap on training wordings.}
\end{table}

\textbf{The tuned compiler is on par with the strongest baselines, and ahead of the parser on rewordings.}
Table~\ref{tab:baselines} suggests that the compiler's advantage lies in \emph{parsing} the intent, not in choosing the knobs.  Given the correct canonical intent, a regression head on frozen embeddings comes close to the joint ceiling (0.29 against 0.28), so it chooses the free knobs about as well as the compiler. On rewordings, however, the head's free knobs follow the embedding of an unseen wording, and its gap rises to 0.41, against 0.37--0.38 for the replay compiler and 0.33 for the joint ceiling, which parse the text themselves. Furthermore, inference-time search replaces the compiler only at a cost of sixty rollouts per intent: even with the correct canonical intent, it needs all sixty to match what the compiler achieves with none; started instead from the compiler's parse with the free knobs at their defaults, and scoring each candidate on three seeds, search stays worse than the compiler at every budget (0.80, 0.71, and 0.62 at ten, thirty, and sixty rollouts). Nevertheless, compilation and search are complementary: warm-starting the same search from the compiler's prediction achieves a gap of 0.21 at sixty rollouts, surpassing both the joint ceiling (0.28) and the lookup (0.31). Finally, the zero-shot backbone fails at the task (0.92, close to the default).

\textbf{Sequential tuning forgets in the executed objective; replay nearly eliminates forgetting.} The third and fourth criteria hold (Table~\ref{tab:exp}). When the adapter is tuned stage after stage with no protection, it forgets on both streams, and far more on Stream~II, where the forgetting measure $F$ is about three times that on Stream~I. On Stream~I, its average gap after the final stage (0.51--0.53) is no better than retrieval (0.41), which copies the label of the most similar training example. Output inspection shows two kinds of error on earlier families: the model drops canonical clauses and invents out-of-schema knob names. Both are small changes to the output text; only the rollout reveals their cost, as a missing clause or knob can change what the controller does. Replay limits forgetting to $F \le 0.07$ across all entries, with AG within 0.05 of the joint ceiling.

\section{Conclusion and future work}
\label{sec:concl}

Our results suggest that large language models can manage continuum platforms at a high level without occupying the fast loop. Formulated as a continually instruction-tuned intent compiler coupled with a deterministic tactical controller, our architecture is auditable and lightweight: an intent is stated in natural language, a compact model compiles it once, off the per-epoch path, and the tactical controller executes it, with a validation gate between the two so that no output of the model reaches the platform without being checked.
Our preliminary findings indicate that catastrophic forgetting is pronounced in executed objectives, that experience replay largely remedies it at the cost of memory overhead for stored exemplars, and that a tuned compact compiler performs on par with simpler translators while being more robust than a grammar parser to rephrased intents.

\textbf{Limitations.} Our preliminary evaluation is small-scale and simulation only, relying on a synthetic intent grammar. The supervision labels are heuristic (best-found via search), not globally optimal. Furthermore, the experiments are limited to a single 2B backbone, two stage orderings, and experience replay as the only anti-forgetting mechanism reported here.

\textbf{Future work.} First, investigating replay-free continual instruction tuning would enable deployment on platforms that are privacy-sensitive or resource-constrained and prohibited from retaining past data. Second, testing larger LLM backbones will clarify whether the compact model's advantages in intent parsing persist at scale, and will quantify the resulting trade-offs in inference latency and edge deployability. Finally, transitioning to a physical testbed platform represents the most significant shift. By replacing simulated rollouts with measured outcomes, it also offers a cheaper source of reference labels: the traces of a manually tuned controller record which parameters served which intents in which contexts, and could replace the search if they prove good enough references.

\begin{acks}
This research was partially supported by the FIS2 Grant from the Italian Ministry of University and Research (Grant ID: FIS2023-03382).
We acknowledge ISCRA for awarding this project access to the LEONARDO supercomputer, owned by the EuroHPC Joint Undertaking, hosted by CINECA (Italy).

\end{acks}

\bibliographystyle{ACM-Reference-Format}
\bibliography{references}

\end{document}